\documentclass[aps,twocolumn,superscriptaddress,nofootinbib]{revtex4-1}

\usepackage{dcolumn}    
\usepackage{pstricks}
\usepackage{color}
\usepackage{graphicx}
\usepackage{amsmath}
\usepackage{mathtools}
\usepackage[T1]{fontenc}
\usepackage[utf8]{inputenc}
\usepackage{amssymb}
\usepackage[colorlinks=false, pdfborder={0 0 0}]{hyperref}

\makeatletter
\newcommand\footnoteref[1]{\protected@xdef\@thefnmark{\ref{#1}}\@footnotemark}
\makeatother

\usepackage{booktabs}

\def\MGUT{M_\mathrm{GUT}}

\def\MI{M_\mathrm{I}}

\def\SO10{\text{SO}(10)}
\newcommand{\G}[1]{\mathcal{G}_\text{#1}}
\def\SU{\,\text{SU}}
\def\GeV{\,\mathrm{GeV}}
\newcommand{\rep}[1]{\mathbf{#1}}
\newcommand{\repb}[1]{\mathbf{\overline{#1}}}

\begin{document}

\title{Realizing unification in two different SO(10) models with one intermediate breaking scale}

\author{Tommy Ohlsson}
\email{tohlsson@kth.se}
\affiliation{Department of Physics,
	School of Engineering Sciences,
	KTH Royal Institute of Technology,
	AlbaNova University Center,
	Roslagstullsbacken 21,
	SE--106 91 Stockholm,
	Sweden}
\affiliation{The Oskar Klein Centre for Cosmoparticle Physics,
	AlbaNova University Center,
	Roslagstullsbacken 21,
	SE--106 91 Stockholm,
	Sweden}

\author{Marcus Pernow}
\email{pernow@kth.se}
\affiliation{Department of Physics,
	School of Engineering Sciences,
	KTH Royal Institute of Technology,
	AlbaNova University Center,
	Roslagstullsbacken 21,
	SE--106 91 Stockholm,
	Sweden}
\affiliation{The Oskar Klein Centre for Cosmoparticle Physics,
	AlbaNova University Center,
	Roslagstullsbacken 21,
	SE--106 91 Stockholm,
	Sweden}

\author{Erik S{\"o}nnerlind}
\email{eriksonn@kth.se}
\affiliation{Department of Physics,
	School of Engineering Sciences,
	KTH Royal Institute of Technology,
	AlbaNova University Center,
	Roslagstullsbacken 21,
	SE--106 91 Stockholm,
	Sweden}
\affiliation{The Oskar Klein Centre for Cosmoparticle Physics,
	AlbaNova University Center,
	Roslagstullsbacken 21,
	SE--106 91 Stockholm,
	Sweden}

\begin{abstract}
We derive the threshold corrections in $\SO10$ grand unified models with the intermediate symmetry being flipped $\SU(5)\times\text{U}(1)$ or $\SU(3)\times\SU(2)\times\text{U}(1)\times\text{U}(1)$, with the masses of the scalar fields set by the survival hypothesis. These models do not achieve gauge coupling unification if the matching conditions do not take threshold corrections into account. We present results showing the required size of threshold corrections for any value of the intermediate and unification scales. In particular, our results demonstrate that both of these models are disfavored since they require large threshold corrections to allow for unification with a predicted proton lifetime above current experimental bounds. 
\end{abstract}

\maketitle

\section{Introduction}
\label{sec:intro}
A popular type of extensions of the Standard Model (SM) are Grand Unified Theories (GUTs)~\cite{Georgi:1974sy}, and especially those based on the $\SO10$ gauge group~\cite{Fritzsch:1974nn}. In order for this type of model to be allowed, it is required that the gauge couplings unify at some high energy scale, defining the scale of unification, and match to the single gauge coupling of the GUT model. This scale has to be high enough so that the predicted rate of proton decay is lower than the limits placed by the non-observation of this phenomenon. 

Symmetry breaking of $\SO10$ down to the SM may proceed in several different chains with multiple intermediate symmetries~\cite{Chang:1984qr}. The renormalization group (RG) running of the gauge couplings varies among these different models, which leads to some being disfavored due to either a too short prediction for the proton lifetime or a failure to unify~\cite{Deshpande:1992au}. However, this conclusion may be modified by including threshold corrections~\cite{Weinberg:1980wa,Hall:1980kf}, which are loop corrections to the matching conditions among gauge couplings imposed at the scale of symmetry breaking. This can have the effect of either saving models that predict a too short proton lifetime~\cite{Parida:1989an,Rani:1993pp,Mohapatra:1992dx,Lee:1994vp,Bertolini:2009qj,Babu:2015bna,Parida:2016hln,Chakraborty:2019uxk,Chakrabortty:2019fov,Meloni:2019jcf} or enabling unification in models that do not unify with tree-level matching conditions~\cite{Lavoura:1993su,Ellis:2015jwa,Schwichtenberg:2018cka,Meloni:2019jcf}.

In this work, we investigate how threshold corrections may enable gauge coupling unification in non-supersymmetric models in which the $\SO10$ symmetry is broken via one intermediate gauge group. Specifically, we focus on two models in which the symmetry breaking proceeds through $\G{51}\equiv\SU(5)\times\mathrm{U}(1)$ with a flipped hypercharge embedding and $\G{3211}\equiv\SU(3)\times\SU(2)\times\text{U}(1)\times\text{U}(1)$. These two models were mentioned, but not investigated in Ref.~\cite{Meloni:2019jcf}, for which threshold corrections were computed for all other $\SO10$ models with zero or one intermediate symmetry. Although the models have been investigated previously in the literature~\cite{Hagiwara:1992ys,Ellis:1995at,Ellis:2002vk,Bertolini:2009qj,LalAwasthi:2011aa,Hirsch:2012kv}, we focus on the numerical computations of the size of the threshold corrections required to achieve gauge coupling unification with an experimentally allowed prediction of the proton lifetime. To this end, we compute the RG running to two-loop order and the threshold corrections to one-loop order. We further consider the effect of kinetic mixing in the $\G{3211}$ model on the size of the required threshold corrections and find that the effect is small.

In Sec.~\ref{sec:rge_thr}, we describe the RG running of gauge couplings and threshold corrections. Then, in Sec.~\ref{sec:51}, we briefly present the model based on the $\G{51}$ symmetry and describe the results for the required threshold corrections in this model. Next, in Sec.~\ref{sec:3211}, we present the same for the model based on the $\G{3211}$ symmetry, including the effect of kinetic mixing. Finally, in Sec.~\ref{sec:conclusions}, we summarize our findings and conclude.

\section{Renormalization group running, threshold effects, and proton decay}
\label{sec:rge_thr}
A prerequisite of grand unification is that the gauge couplings in the low-energy model emerge from the one corresponding to the GUT symmetry, in our case $\SO10$. This is reconciled with their differing values at low energy by the RG running, through which the gauge couplings vary with the energy scale. To two-loop order in perturbation theory, the RG equations read
\begin{equation}
\frac{\mathrm{d}\alpha_i^{-1}(\mu)}{\mathrm{d}\ln\mu} = -\frac{a_i}{2\pi} - \sum_j \frac{b_{ij}}{8\pi^2 \alpha_j^{-1}(\mu)},
\end{equation}
where $\alpha_i \equiv g_i^2/(4\pi)$, with $g_i$ being the gauge coupling corresponding to the $i$th gauge group in the model, and $a_i$ and $b_{ij}$ are the one- and two-loop coefficients, respectively~\cite{Jones:1981we,Machacek:1983tz}. The measured values of the gauge couplings at the electroweak scale $M_\text{Z}\simeq 91.1876\,\text{GeV}$ are~\cite{PDG}
\begin{equation}
(\alpha_\text{3}^{-1}(M_\text{Z}), \alpha_\text{2}^{-1}(M_\text{Z}), \alpha_\text{1}^{-1}(M_\text{Z})) \simeq (8.50, 29.6, 59.0).
\end{equation}
These constitute the boundary conditions of the system of equations given by the RGEs. 

At a scale $M_{m\rightarrow n}$ of symmetry breaking of a group $\mathcal{G}_m$ to another group $\mathcal{G}_n$, the matching conditions for the gauge couplings in the lower- and higher-energy theories including threshold corrections $\lambda^m_n$ are 
\begin{equation}\label{eq:matching}
\alpha_n^{-1}(M_{m\rightarrow n}) = \alpha_m^{-1}(M_{m\rightarrow n}) - \frac{\lambda^m_n}{12\pi}.
\end{equation}
To one-loop order, the threshold corrections $\lambda^m_n$ are given by~\cite{Hall:1980kf,Weinberg:1980wa}
\begin{align}\label{eq:thresholds}
\lambda^m_n = &\sum_{i\, \in\, \text{vectors}}k_{V_i} S_2(V_i)\nonumber \\
&+ \sum_{i\, \in\, \text{scalars}} \kappa_{S_i} k_{S_i} S_2(S_i) \ln\left( \frac{M_{S_i}}{M_{m\rightarrow n}}\right),
\end{align}
where $\kappa_{S_i}$ are $1$ or $2$ for real or complex representations, while $k_{V_i}$ and $k_{S_i}$ are the multiplicities of the vector and scalar fields $V_i$ and $S_i$, respectively, $S_2(r)$ is the Dynkin index of the representation $r$, and $M_{S_i}$ are the masses of $S_i$. It is assumed that all superheavy gauge boson masses lie at the symmetry breaking scale such that they do not have any mass-dependent contributions and that there are no additional fermions that contribute to the threshold corrections. For readability, we let $\eta_i \equiv \ln (M_{S_i}/M_{m\rightarrow n})$.

The allowed scale $\MGUT$ is a relevant prediction, since it is related to proton decay. It gives the prediction for the proton lifetime in the most constraining channel to be~\cite{Nath:2006ut,Babu:2010ej}
\begin{equation}
\tau(p\rightarrow e^+\pi^0)\simeq (7.47\times 10^{35}\, \text{yr}) \left(\frac{\MGUT}{10^{16}\GeV}\right)^4 \left(\frac{0.03}{\alpha_\text{GUT}}\right)^2,
\end{equation}
where $\alpha_\text{GUT}$ is the gauge coupling at $\MGUT$. This prediction is to be compared with the lower bound stemming from the non-observation of proton decay. The currently best bound comes from Super-Kamiokande~\cite{Mine:2016mxy,Miura:2016krn,Abe:2014mwa,Abe:2013lua} which gives $\tau(p\rightarrow e^+\pi^0)>1.67\times 10^{34}\,\text{yr}$ at $90~\%$ confidence level.

It should be noted that models based on a broken $\SO10$ symmetry in general exhibit topological defects that place constraints on the models. We do not investigate this further and instead refer the reader to Refs.~\cite{Lazarides:1980cc,Lazarides:1980va,Lazarides:1981fv,Kibble:1982ae,Kibble:1982dd,Weinberg:1983bf,Bhattacharjee:1991zm,Davis:1994py,Chakrabortty:2017mgi,Chakrabortty:2019fov,Lazarides:2019xai}.

\section{Flipped SU(5)$\times$U(1)}
\label{sec:51}
In the flipped $\SU(5)\times\text{U}(1)_{X}$ model, the SM hypercharge is a linear combination of the Abelian charge from within $\SU(5)$ and the external $\text{U}(1)_{X}$ charge. Models of this kind have been previously considered in \textit{e.g.} Refs.~\cite{DeRujula:1980qc,Barr:1981qv,Derendinger:1983aj,Antoniadis:1987dx,Ellis:1988tx,Hagiwara:1992ys,Ellis:1995at,Ellis:2002vk}. The reason why the flipped hypercharge embedding is considered is that the standard embedding would only move the problem of unification from the $\SO10$ breaking scale to the $\SU(5)$ breaking scale.

Apart from the fermions in the $\rep{16}_\text{F}$ representation of $\SO10$, the model contains scalars in the $\rep{10}_\text{H}$, $\rep{45}_\text{H}$, and $\repb{126}_\text{H}$ representations\footnote{Different particle content is also possible. Since we are not concerned with the details of model building, we simply take the particle content required to achieve the breaking as well as provide two independent Yukawa couplings.} of $\SO10$, resulting in 29 scalar masses. The masses of the various scalar components are assumed to follow the survival hypothesis~\cite{Georgi:1979md,delAguila:1980qag,Mohapatra:1982aq}.

The matching conditions at $\MI$ are such that the gauge couplings corresponding to $\SU(2)_\text{L}$ and $\SU(3)_\text{C}$ in the SM both match to the one corresponding to $\SU(5)$. Therefore, the matching conditions at $\MI$ with threshold conditions read
\begin{align}
\alpha_{X}^{-1}(\MI) &=\frac{1}{24} \Big[25 \alpha_\text{1}^{-1}(\MI) - \alpha_\text{2}^{-1}(\MI) \nonumber\\
&\quad + \frac{1}{12\pi}(25\lambda_\text{1}^{51} - \lambda_\text{2}^{51})\Big],\\
 \alpha_\text{3}^{-1}(\MI) &- \alpha_\text{2}^{-1}(\MI) + \frac{1}{12\pi}(\lambda^{51}_\text{3} - \lambda^{51}_\text{2}) = 0.\label{eq:32_matching}
\end{align}
Similarly, we can write the matching condition at $\MGUT$ as 
\begin{equation}
\alpha_5^{-1}(\MGUT) - \alpha_{X}^{-1}(\MGUT) + \frac{1}{12\pi}(\lambda_5^{10} - \lambda_{X}^{10}) = 0.
\end{equation}

\begin{figure}
\includegraphics[width=\columnwidth]{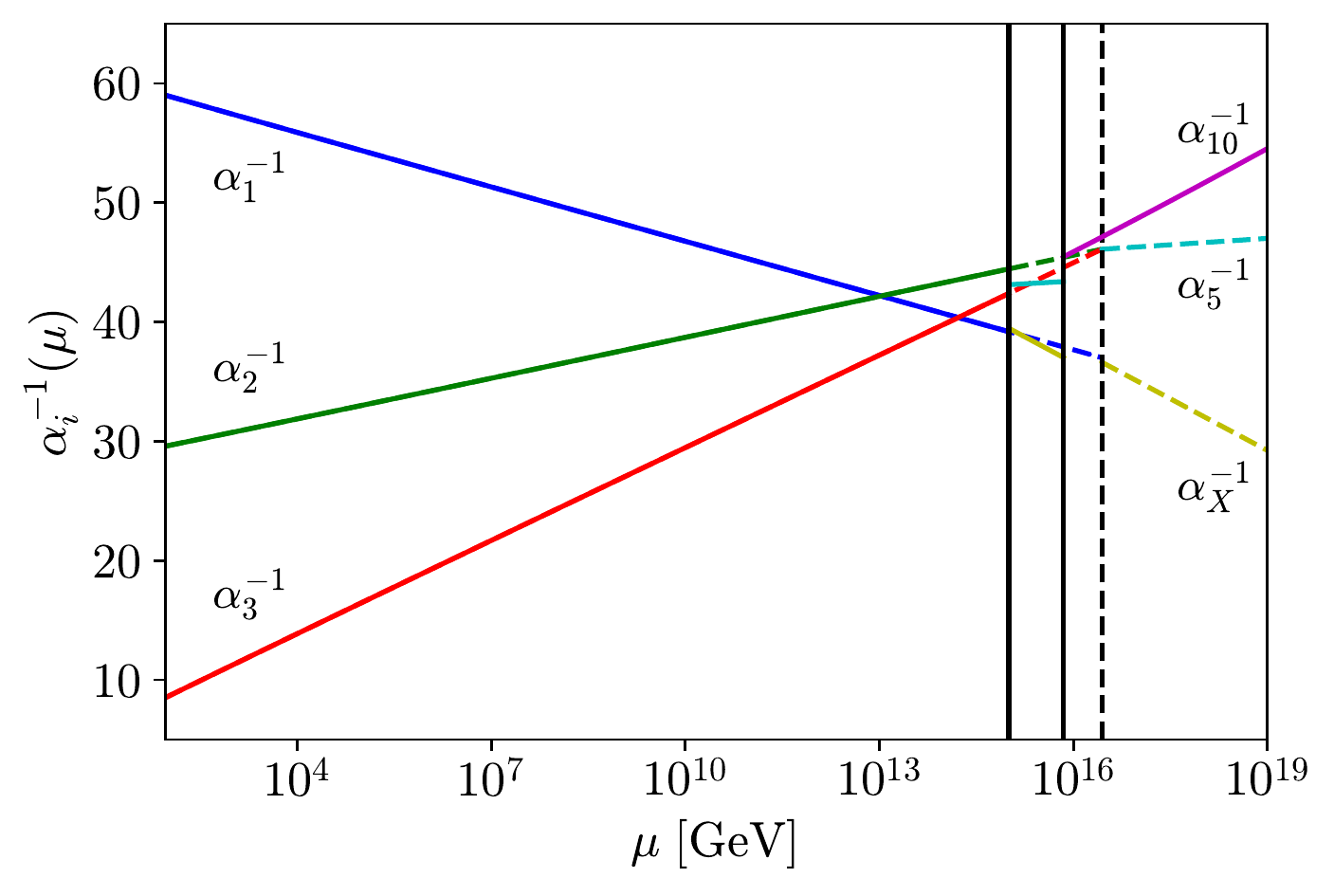}
\caption{\label{fig:51Running}RG running of gauge couplings $\alpha_i^{-1}$ as a function of the energy scale $\mu$ in the $\G{51}$ model. Dashed lines indicate the case of no threshold corrections. Solid lines show a representative example of gauge coupling unification taking into account threshold corrections, with $\MI\simeq10^{15}\,\text{GeV}$ and $\MGUT\simeq7\times10^{15}\,\text{GeV}$ and with $\eta_i\in[-4,+4]$.}
\end{figure}

As can be seen by the dashed lines in Fig.~\ref{fig:51Running}, unification cannot be achieved in this model without sufficiently large threshold corrections. The reason for this is that $\MI$ is set by the intersection of $\alpha_\text{2}^{-1}$ and $\alpha_\text{3}^{-1}$, as demonstrated by Eq.~\eqref{eq:32_matching} with threshold corrections set to zero. We then require the threshold corrections $\lambda_\text{3}^{51}$ and $\lambda_\text{2}^{51}$ to be such that $\MI$ is pushed to a lower value. Since the two gauge couplings in $\G{51}$ diverge, $\MI$ should preferably be below the intersection of $\alpha_\text{2}^{-1}$ and $\alpha_\text{1}^{-1}$, or close enough to it such that $\alpha_{X}^{-1}$ and $\alpha_5^{-1}$ can unify at $\MGUT$ without excessively large threshold corrections.

\begin{figure}
\includegraphics[width=\columnwidth]{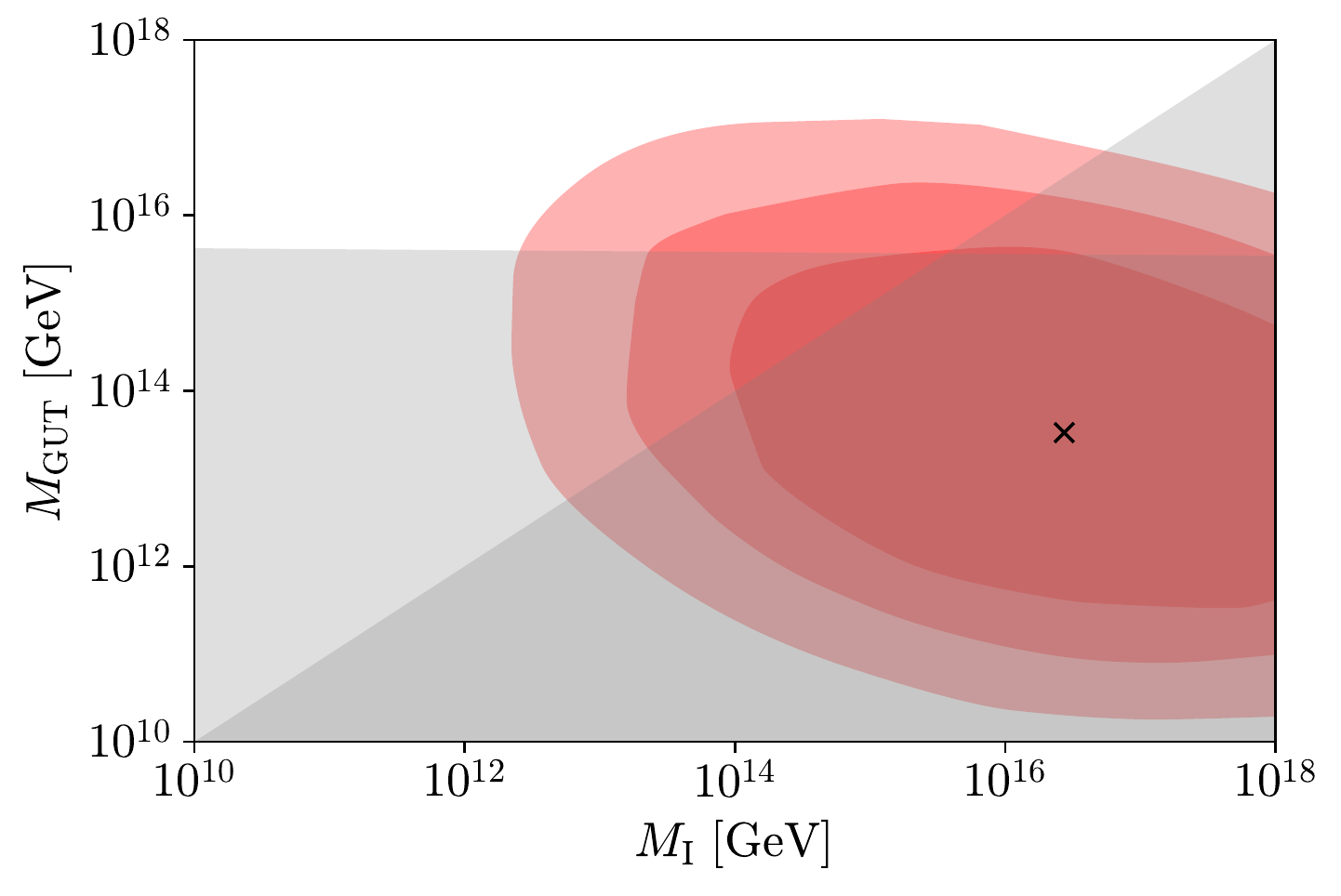}
\caption{\label{fig:51}Possible scales $\MI$ and $\MGUT$ with successful gauge coupling unification due to threshold corrections in the $\G{51}$ model. Red regions denote solutions with $\eta_i\in[-3,+3]$ (dark red), $[-4,+4]$ (red), and $[-5,+5]$ (light red) and the ``$\times$'' denotes the solution without threshold corrections. The horizontal and slanted gray shaded regions are ruled out by a too short proton lifetime and $\MI>\MGUT$, respectively.}
\end{figure}

To illustrate the typical size of the perturbations of scalar masses $\eta_i$ required in order to allow for unification, we randomly sample each of the 29 $\eta_i$ parameters uniformly in the ranges $[-3,+3]$, $[-4,+4]$, and $[-5,+5]$ and compute the scales $\MI$ and $\MGUT$ for each sampled point. During this sampling process, we allow all the $\eta_i$ parameters to vary independently and simultaneously. The points that corresponded to gauge coupling unification are plotted in Fig.~\ref{fig:51}, in which the \texttt{ConvexHull} method from the \texttt{SciPy} package~\cite{2020SciPy-NMeth} is used to illustrate the regions. Only points lying above the slanted gray (corresponding to the unphysical region $\MGUT < \MI$) and the nearly horizontal region (corresponding to a too short proton lifetime) are allowed.

As can be observed, only a few points with $\eta_i\in[-3,+3]$ are allowed. This illustrates that the model requires either fine-tuned or large threshold corrections in order to achieve unification. Both of these cases can be considered quite unnatural.

Since the $\G{51}$ model contains leptoquark gauge bosons, one has to take into account proton decay also in the intermediate symmetry, with a dependence on the scale $\MI$. Following Ref.~\cite{Ellis:2020qad}, we find that values of $\MI\gtrsim 10^{15}\,\text{GeV}$ evade the bound on proton lifetime, both in the $p\rightarrow e^+\pi^0$ and $p\rightarrow \bar\nu \pi^+$ channels. This further reduces the allowed parameter space in Fig.~\ref{fig:51}, but still allows the points with $\eta_i\in[-3,+3]$. However, as discussed in Ref.~\cite{Barr:2013gca}, these predictions can be largely altered by the values of the fermion mixing angles at that scale.

An example of unification is shown by the solid lines in Fig.~\ref{fig:51Running}. Here, we choose $\eta_i\in[-4,+4]$ such that $\MI\simeq 10^{15}\,\text{GeV}$ and $\MGUT\simeq 7\times10^{15}\,\text{GeV}$. Although the lines do not meet, the threshold corrections are such that the matching produces a unique value of the $\SO10$ gauge coupling, thus signifying unification.

\section{SU(3)$\times$SU(2)$\times$U(1)$\times$U(1)}
\label{sec:3211}

The other model we consider is based on the $\SU(3)_\text{C}\times\SU(2)_\text{L}\times\text{U}(1)_{R}\times\text{U}(1)_{B-L}$ symmetry~\cite{Chang:1984uy,Parida:1989an,Parida:1991sj,Bertolini:2009qj,LalAwasthi:2011aa,Hirsch:2012kv}. We take the scalar sector to contain the $\rep{210}_\text{H}$, $\rep{10}_\text{H}$, and $\repb{126}_\text{H}$ representations\footnote{As in Sec.~\ref{sec:51}, different particle content is also possible. Since we are not concerned with the details of model building, we simply take the particle content required to achieve the breaking as well as provide two independent Yukawa couplings. In particular, a $\rep{45}_\text{H}$ could be used instead of $\rep{210}_\text{H}$, but that would require two separate vevs from within the $\rep{45}_\text{H}$. We expect this to require somewhat larger values of $\eta_i$ as there are fewer individual masses that can be perturbed around the symmetry breaking scale.} with the masses set by the survival hypothesis. This results in 65 different scalar masses in this model. As in Sec.~\ref{sec:51}, we have the fermions in the $\rep{16}_\text{F}$. 

When matching the SM gauge couplings to those of $\G{3211}$, the SM hypercharge becomes a linear combination of the two Abelian charges of $\G{3211}$. Therefore, there is no unique matching when solving the problem from low to high energy. For this reason, we introduce an auxiliary parameter $x$ that parametrizes the relation between the two Abelian gauge couplings in $\G{3211}$ at $\MI$, \textit{i.e.} $\alpha_{B-L}^{-1}(\MI) = x \alpha_{R}^{-1}(\MI)$. With this, the matching conditions at $\MI$ with threshold corrections read
\begin{align}
\alpha_{B-L}^{-1}(\MI) &= x \left(\frac25 x + \frac35 \right)^{-1} \left[\alpha_\text{1}^{-1}(\MI)+ \frac{\lambda_\text{1}^{3211}}{12\pi}\right],\label{eq:match_3221_SM_1}\\
\alpha_{R}^{-1}(\MI) &= \left(\frac25 x + \frac35 \right)^{-1} \left[\alpha_\text{1}^{-1}(\MI)+ \frac{\lambda_\text{1}^{3211}}{12\pi}\right],\label{eq:match_3221_SM_2}\\
\alpha_\text{2L}^{-1}(\MI) &= \alpha_\text{2}^{-1}(\MI) +\frac{\lambda^{3211}_\text{2}}{12\pi},\label{eq:match_3221_SM_3}\\
\alpha_\text{3C}^{-1}(\MI) &=\alpha_\text{3}^{-1}(\MI)+\frac{\lambda^{3211}_\text{3}}{12\pi}.\label{eq:match_3221_SM_4}
\end{align}
At $\MGUT$, all gauge couplings must match to the one corresponding to $\SO10$. There are then three independent matching conditions given by
\begin{align}
\alpha_{B-L}^{-1}(\MGUT) - \alpha_{R}^{-1}(\MGUT) + \frac{1}{12\pi}(\lambda_{B-L}^{10} - \lambda_{R}^{10}) &= 0,\label{eq:match_SO10_3221_1}\\
\alpha_{R}^{-1}(\MGUT) - \alpha_\text{2L}^{-1}(\MGUT) + \frac{1}{12\pi}(\lambda_{R}^{10} - \lambda_\text{2L}^{10}) &= 0,\label{eq:match_SO10_3221_2}\\
\alpha_\text{2L}^{-1}(\MGUT) - \alpha_\text{3C}^{-1}(\MGUT) + \frac{1}{12\pi}(\lambda_\text{2L}^{10} - \lambda_\text{3C}^{10}) &= 0.\label{eq:match_SO10_3221_3}
\end{align}

\begin{figure}
\includegraphics[width=\columnwidth]{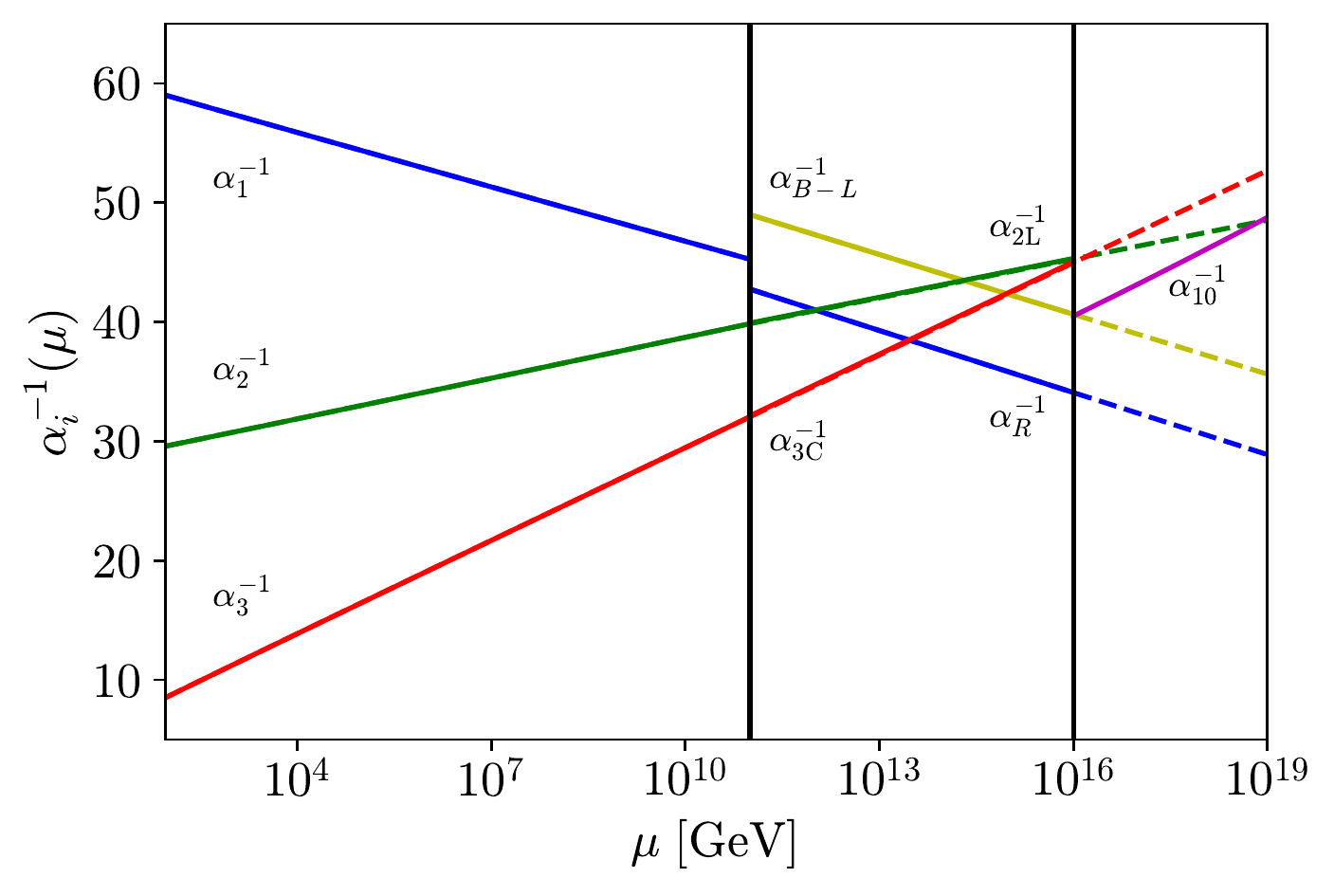}
\caption{\label{fig:3211Running}RG running of gauge couplings $\alpha_i^{-1}$ as a function of the energy scale $\mu$ in the $\G{3211}$ model. Dashed lines indicate the case of no threshold corrections. Solid lines show a representative example of gauge coupling unification taking into account threshold corrections,  with $\MI=10^{11}\,\text{GeV}$ and $\MGUT=10^{16}\,\text{GeV}$ and with $\eta_\text{max}\simeq4.2$.}
\end{figure}

As illustrated by the example shown in Fig.~\ref{fig:3211Running}, unification is not possible without threshold corrections in this model. In order to numerically investigate the effect of threshold corrections in the model, we first note that the threshold corrections at $\MI$ are negligible compared to those at $\MGUT$. The reason is that very few fields lie around $\MI$ in this model, since the $\G{3211}$ symmetry is broken down to SM by a vacuum expectation value in the $(\rep{1},\rep{1})_{1,2}$ representation. The numerical computations are therefore performed by neglecting threshold corrections at $\MI$, and then checking that they indeed give rise to negligible corrections to the derived scales. 

The threshold corrections then enter only through the three differences $\lambda_{B-L}^{10} - \lambda_{R}^{10}$, $\lambda_{R}^{10} - \lambda_\text{2L}^{10}$, and $\lambda_\text{2L}^{10} - \lambda_\text{3C}^{10}$ in Eqs.~\eqref{eq:match_SO10_3221_1}--\eqref{eq:match_SO10_3221_3}. For each set of scales $\MI$ and $\MGUT$ and auxiliary parameter $x$, we can find the required values of these differences that allow the matching of the four gauge couplings in $\G{3211}$ to $\SO10$. Once these are found, we look for values of $\eta_i$ that can result in these differences. This can be done by expressing the problem as a system of linear equations to be solved for the vector of $\eta_i$. The vector that has the smallest Euclidean norm is then found by the \texttt{lstsq} method in the \texttt{NumPy} package~\cite{harris2020array}. This solution is then labeled by its maximum value $\eta_\text{max}$. Finally, to remove the dependence on the auxiliary variable $x$, we repeat this whole procedure to find the value of $x$ for each set of scales $\MI$ and $\MGUT$ that minimizes the value of $\eta_\text{max}$.

\begin{figure}
\includegraphics[width=\columnwidth]{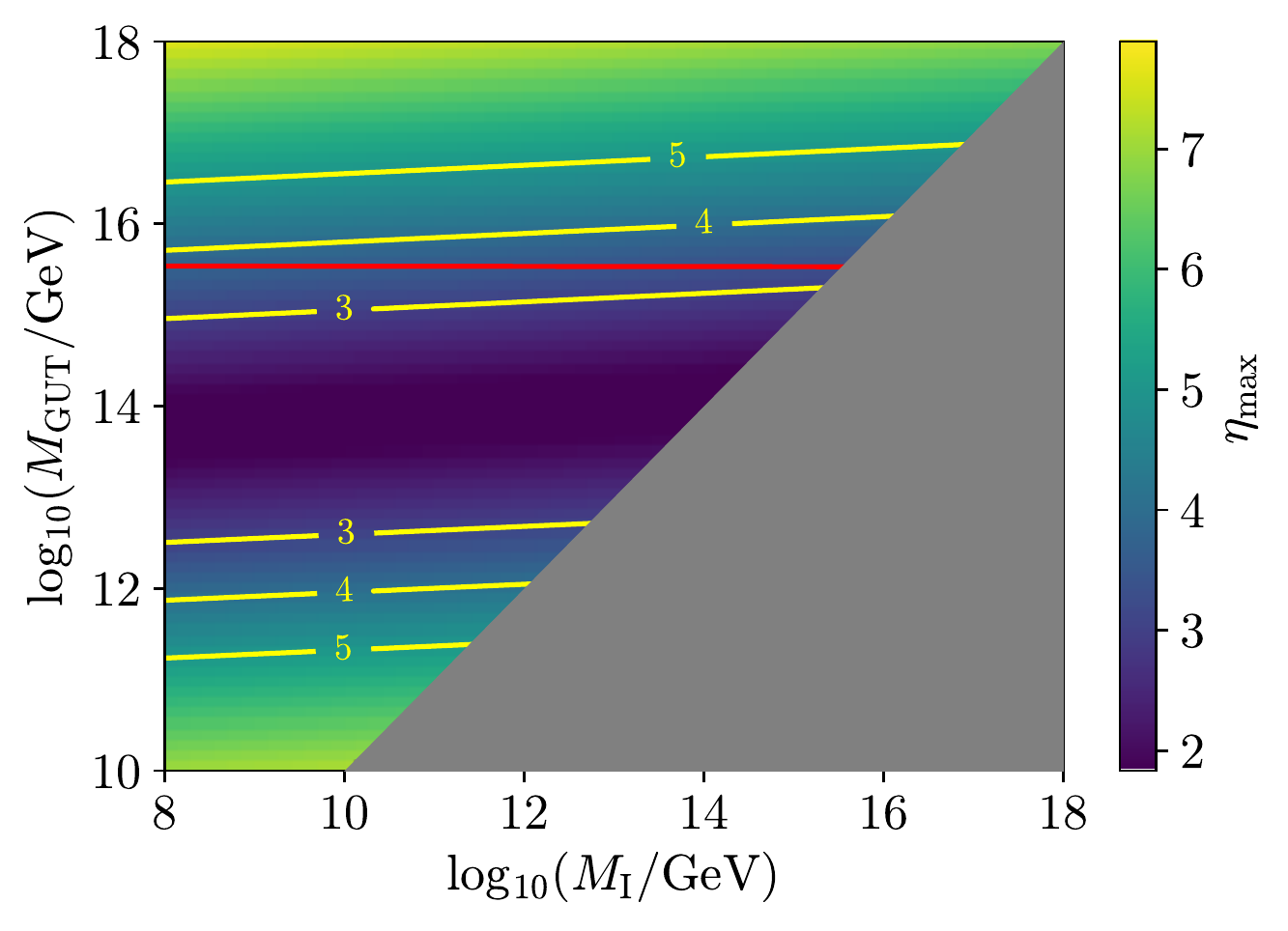}
\caption{\label{fig:3211}Possible scales $\MI$ and $\MGUT$ with successful gauge coupling unification due to threshold corrections in the $\G{3211}$ model. The color corresponds to the value of $\eta_\text{max}$ required to achieve the corresponding scales. Points below the horizontal red line are ruled out by a too short proton lifetime and the slanted gray shaded region is forbidden by $\MI>\MGUT$.}
\end{figure}

The results of this procedure is displayed in Fig.~\ref{fig:3211}, in which the smallest possible value of $\eta_\text{max}$ is shown for each combination of scales $\MI$ and $\MGUT$. The gray shaded region corresponds to the unphysical region of $\MGUT < \MI$ and the red line is the current limit set by the absence of proton decay. In comparing with the dashed lines in Fig.~\ref{fig:3211Running}, we find that, with $\MI=10^{11}\,\text{GeV}$, the value of $\MGUT$ that requires the smallest threshold corrections corresponds to a value that lies inside the quadrilateral formed by the four gauge couplings. However, this is not allowed by the bounds on the proton lifetime. The smallest value of $\eta_\text{max}$ that is allowed is about $3.34$, corresponding to $\MI$ and $\MGUT$ being very close, as shown in Fig.~\ref{fig:3211}.

For the sake of completeness, we also investigate how kinetic mixing between the Abelian groups~\cite{Holdom:1985ag,delAguila:1988jz} changes the result. This introduces an off-diagonal gauge coupling, whose RG running affects that of the other gauge couplings, calculated using the \texttt{PyR@TE} package~\cite{Lyonnet2014,Lyonnet:2016qyu,Lyonnet:2016xiz}. It further affects the matching conditions Eqs.~\eqref{eq:match_3221_SM_1}--\eqref{eq:match_3221_SM_4} at $\MI$, given \textit{e.g.} in Refs.~\cite{Bertolini:2009qj,Bertolini:2013vta,Fonseca:2013bua}. Finally, at $\MGUT$, we impose the condition that the off-diagonal coupling is zero, since the mixing is not generated by the symmetry breaking of $\SO10$ but rather from the RG running~\cite{delAguila:1988jz,Rizzo:1998ut}. 

\begin{figure}
\includegraphics[width=\columnwidth]{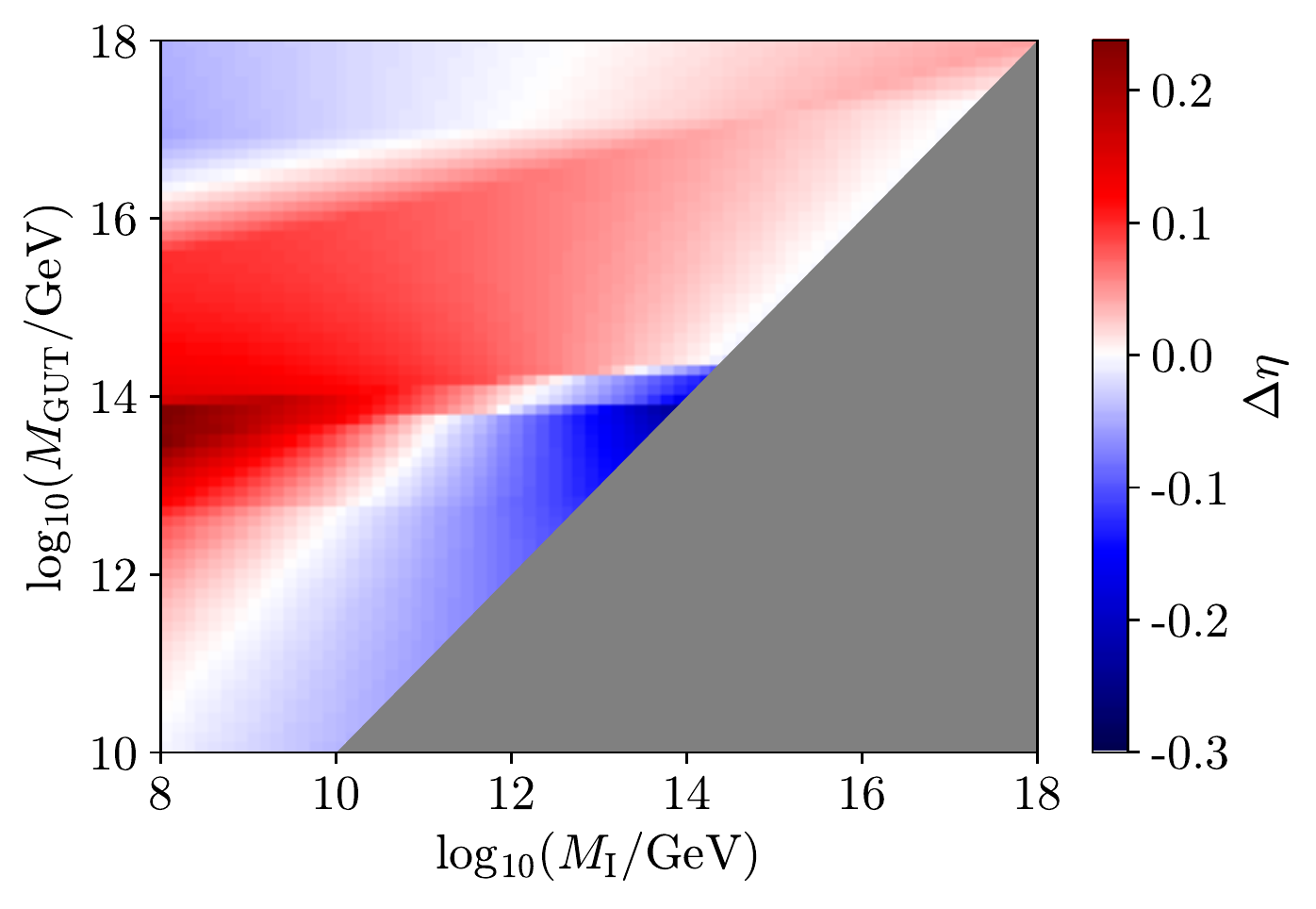}
\caption{\label{fig:3211_mixing}Relative changes in $\eta_\text{max}$ due to kinetic mixing of the Abelian gauge couplings in the $\G{3211}$ model in the \mbox{$\MI$--$\MGUT$} plane. Red (blue) regions indicate that smaller (larger) threshold corrections are needed than for the case without kinetic mixing. The gray region is ruled out by $\MI>\MGUT$.}
\end{figure}

We then perform an analysis similar to the one described above, but with the additional inclusion of kinetic mixing. This leads to slightly different values of $\eta_\text{max}$, from which we calculate the relative difference $\Delta \eta$. This is shown in Fig.~\ref{fig:3211_mixing}, which shows that the largest relative difference found is about 20~\%. However, in the region that is allowed by the proton lifetime bounds, the relative difference $\Delta\eta$ is less than about 9.7~\%.

\section{Summary and Conclusions}
\label{sec:conclusions}
We have presented computations of the extent to which threshold corrections can allow for gauge coupling unification in two $\SO10$ models with $\SU(5)\times\text{U}(1)$ and $\SU(3)\times\SU(2)\times\SU(2)\times\text{U}(1)$ as intermediate symmetry groups. Although other particle contents are possible for these models, the results presented in this work are representative for these types of models. The RG equations have been solved numerically at two-loop level and the matching conditions among the gauge couplings at symmetry breaking scales include threshold corrections at one-loop level. Furthermore, we have investigated the effect of kinetic mixing on the results in the $\SU(3)\times\SU(2)\times\SU(2)\times\text{U}(1)$ model.

In both of these models, our results have suggested that threshold corrections with $\eta_i\in[-4,+4]$ are required to achieve unification while evading the bound from proton decay. Slightly smaller values of $\eta_i$ are possible, but only for very small regions in parameter space, suggesting that these scenarios require fine-tuning of the scalar masses. With such large threshold corrections, the naturalness of such theories can be questioned. However, given that GUTs already suffer from naturalness problems due to the large separation of scales, the required threshold correction may be a minor issue. Furthermore, light states may arise naturally due to the presence of pseudo-Goldstone bosons~ \cite{Bertolini:2012im}. Finally, the effects of kinetic mixing have been shown to have a less than 10~\% effect on the required threshold corrections.

This investigation has neglected any constraints from the scalar potential. For example, there can exist correlations among the masses of the different scalar fields which restrict the allowed threshold corrections. This would be particularly interesting in the $\SU(5)\times\text{U}(1)$ model, since that is the minimal $\SO10$ model allowed without a tachyonic scalar spectrum~\cite{Yasue:1980fy} or the need for loop-level corrections to the scalar potential~\cite{Bertolini:2009es}. Thus, it would be of interest to investigate whether constraints from the scalar potential are compatible with the threshold corrections that allow the model to evade the proton lifetime bound. Furthermore, the survival hypothesis was imposed for reasons of simplicity and is not a strict requirement. Any departure from it could affect the results of this investigation. For large values of $\MGUT$, it is also possible that Planck-suppressed higher-order operators affect the result by both modifying the matching conditions~\cite{Shafi:1983gz,Hill:1983xh,Calmet:2008df} and the proton decay rate~\cite{Harnik:2004yp,Barr:2012xb}. Finally, we have also not taken into account any phenomenology relating to the value of $\MI$, which can place further constraints on the viability of the models. These constraints can, for example, stem from neutrino masses and leptogenesis.

\begin{acknowledgments}
We would like to thank Davide Meloni for useful discussions. T.O.~acknowledges support by the Swedish Research Council (Vetenskapsr\r{a}det) through contract No.~2017-03934. Numerical computations were performed on resources provided by the Swedish National Infrastructure for Computing (SNIC) at PDC Center for High Performance Computing (PDC-HPC) at KTH Royal Institute of Technology in Stockholm, Sweden under project number SNIC 2020/5-122.
\end{acknowledgments}

\end{document}